\documentclass{aa}

\newcommand{\gtappeq}{\raisebox{-0.6ex}{$\,\stackrel
{\raisebox{-.2ex}{$\textstyle >$}}{\sim}\,$}}

\newcommand{\ltappeq}{\raisebox{-0.6ex}{$\,\stackrel
{\raisebox{-.2ex}{$\textstyle <$}}{\sim}\,$}}

\begin{document}

\thesaurus{06 % Formation, Structure and Evolution of Stars
(
02.01.2 %Accretion, accretion disks
08.02.1 %binaries, close
08.14.2 %novae, CVs
08.09.2 %Stars: individual -- T Pyx
08.13.2 %Stars: mass-loss
08.19.4 %supernovae: general
)
}
%might think about x-rays as a keyword

\title{Assisted stellar suicide: the wind-driven evolution of 
the recurrent nova T~Pyxidis}

\author{
Christian Knigge
\inst{1,2}
\fnmsep
\thanks{Hubble Fellow}
\and
Andrew R. King
\inst{3}
\and
Joseph Patterson
\inst{2}}

\institute{
Department of Physics and Astronomy,
University of Southampton,
Highfield,
Southampton SO17 1BJ, UK
\and
Department of Astronomy,
Columbia University,
550 West 120th Street,
New York, NY 10027, USA 
\and
Astronomy Group, 
University of Leicester, 
Leicester LE1 7RH, UK}

\offprints{C. Knigge}
\mail{christian@astro.soton.ac.uk}

\date{Received 00-00-0000 / Accepted 00-00-0000}

\maketitle

\begin{abstract}

We show that the extremely high luminosity of the short-period
recurrent nova T~Pyx in quiescence can be understood if this system is 
a wind-driven supersoft x-ray source (SSS). In this scenario, a
strong, radiation-induced 
wind is excited from the secondary star and accelerates the binary
evolution. The accretion rate is therefore much higher than in an
ordinary cataclysmic binary at the same orbital period, as is the
luminosity of the white dwarf primary. In the steady state, the
enhanced luminosity is just sufficient to maintain the wind from the
secondary. The accretion rate and luminosity predicted by the
wind-driven model for T~Pyx are in good agreement with the
observational evidence. X-ray observations with Chandra or XMM 
may be able to confirm T~Pyx's status as a SSS. 

T~Pyx's lifetime in the wind-driven state is on the order of a million 
years. Its ultimate fate is not certain, but the system may very well
end up destroying itself, either via the complete evaporation of the
secondary star, or in a Type~Ia supernova if the white dwarf 
reaches the Chandrasekhar limit. Thus either the primary, the secondary, 
or both may currently be committing assisted stellar suicide.


\end{abstract}

\section{Introduction}
\label{introduction}
The recurrent nova T~Pyx is the most prolific of all recurrent 
novae, with a mean outburst recurrence time of only 22 years. But even 
away from outburst, T~Pyx's observational characteristics are 
remarkable. Most importantly, its quiescent bolometric luminosity is 
$L_{\rm bol} \gtappeq 10^{36}$~ergs~s$^{-1}$ (Patterson et al. 1998). This
is much higher than expected for a short-period cataclysmic variable (CV) 
like T~Pyx, whose orbital period of $P_{\rm orb} = 1.8$~hr places it well 
below the period gap.\nocite{patterson1} 

The implied disagreement with theoretical expectations is severe. 
For reasonable estimates of T~Pyx's system parameters, the standard
model of CV evolution predicts mass transfer to be driven by
gravitational radiation at a rate of $\dot{M}_{\rm gr} \simeq 4 
\times 10^{-11}{\rm ~{M}_{\sun}~yr^{-1}}$. 
The corresponding accretion luminosity is only $L\simeq 9 \times
10^{32}$~erg~s$^{-1}$. Thus the gap between T~Pyx's observed
luminosity and the theoretically predicted value spans at least 
three orders of magnitude.

In order to account for T~Pyx's extreme luminosity without having
to postulate a similarly extreme mass-transfer rate for the system, 
it has been suggested that, even in quiescence, the system may
be powered predominantly by nuclear burning on the surface of the
white dwarf (WD), rather than by accretion (e.g. Webbink et al. 1987;
Patterson et al. 1998). However, even this can only reduce the
luminosity gap by about an order of magnitude
(c.f. Section~\ref{basics}). Thus regardless of the actual mode of
radiation energy release, T~Pyx's extreme luminosity implies an
accretion rate $\dot{M}_{\rm acc} \gtappeq (4-40) \times
10^{-9}{\rm~{M_{\sun}}~yr^{-1}}$.\nocite{webbink1} This lower limit is
in line with estimates based on theoretical nova models: in order to 
reproduce T~Pyx's short outburst recurrence time, an accretion rate 
$\dot{M}_{\rm acc} \gtappeq 10^{-8}{\rm  ~M_{\sun}~yr^{-1}}$ appears to be
required (Livio \& Truran 1992; Prialnik \& Kovetz 1995; Contini \&
Prialnik 1997).

Clearly, T~Pyx poses a fundamental challenge to our understanding of 
CV evolution. Put concisely, the question is this: {\em why is 
T~Pyx's accretion rate so abnormally high relative to ``ordinary''
CVs?} The answer we propose in this {\em Letter} is that T~Pyx has
left the standard CV evolutionary track completely and is instead
currently evolving as a wind-driven supersoft X-ray source. 

\section{T~Pyx as a Wind-driven Supersoft X-ray Source}

\subsection{Motivation}

T~Pyx's observational characteristics in quiescence are clearly
extreme among CVs. However, as recently pointed out by Patterson et 
al. (1998) they are actually a good match to the properties of the 
supersoft x-ray sources (SSSs). Despite this empirical connection, 
the standard model for binary SSSs -- thermal-timescale mass transfer
from a companion initially more massive than the WD -- cannot work for 
T Pyx: the shortest orbital period found in the detailed study of this
phase by Deutschmann (1998) is about 6~hr, much longer than T Pyx's
1.8~hr. In fact, King et al. (2000) find that the two shortest-period
systems among known binary SSSs -- RX J0537.7$-$7034 ($P_{\rm orb} \simeq
3.5$~hr) and SMC 13 ($P_{\rm orb} \simeq 4.1$~hr) -- are already difficult
to account for within the thermal-timescale mass-transfer framework,
even allowing for non-conservative evolution.

We therefore propose that T~Pyx is instead a member of an (even
more) exotic class of SSSs -- the {\em wind-driven supersoft
sources}. The existence of this class has recently been suggested by 
van Teeseling \& King (1998, hereafter VK98; see also King \& van
Teeseling 1998, hereafter KV98). These authors show 
that if the secondary star in a close binary system like T~Pyx is
strongly irradiated by soft x-rays, a powerful wind may be driven 
from its surface. Under certain conditions, the mass and angular
momentum loss in this wind can dominate the binary evolution, which 
in turn can drive mass transfer at a rate comparable to the wind 
mass-loss rate. If the wind-driven mass-transfer rate is high enough 
to power the irradiating source, a self-sustaining, stable, wind-driven 
state is created. All that is required in this scenario is an event 
to trigger the wind-driven evolution.
\nocite{teeseling2,king5}

\subsection{Theoretical Background}
\label{basics}

We take as our starting point Equation~7 in VK98, which gives the 
expected mass-loss rate from a low-mass, near-main-sequence secondary 
star irradiated by intense soft x-ray emission from the vicinity of
the primary as 
\begin{equation}
\dot{M}_{w2} \simeq -3 \times 10^{-7} \phi \frac{r_2}{a_{11}} (m_2 L_{37}
\eta_s)^{1/2} {\rm ~M_{\sun} yr^{-1}}.
\label{mdot_wind1}
\end{equation}
In this equation, $m_2=M_2/M_{\sun}$ and $r_2=R_2/R_{\sun}$ are 
the mass and radius of the secondary,
$a_{11}=a/(10^{11}~{\rm cm})$ is the binary separation and 
$L_{37}=L/(10^{37}~{\rm erg~s^{-1}})$ is the luminosity of the
irradiating source. The factor $\eta_s$ measures the
efficiency of the irradiating spectrum in producing the wind, and
$\phi$ measures the area of the mass-losing regions on the
secondary's surface, relative to $\pi R^2_2$.

The irradiating luminosity is most conveniently parameterized in
terms of the luminosity expected for steady shell burning,
$L_{\rm sb}$, which is 
\begin{equation}
L_{\rm sb} = 2.9 \times 10^{37} \left(\frac{\dot{M}_{\rm acc}}{10^{-7}~{\rm
M_{\sun}~yr^{-1}}}\right) {\rm erg~s^{-1}}
\end{equation}
(Iben 1982). Defining $L_{{\rm sb}, 37} = L_{\rm sb}/(10^{37}~{\rm
erg~s^{-1}})$, we therefore write the irradiating luminosity as 
\begin{equation}
L_{37} = \eta_a L_{{\rm sb},37} 
       = 2.9 \eta_a \left(\frac{\dot{M}_{\rm acc}}{10^{-7}~{\rm M_{\sun}~yr^{-1}}}\right) 
\label{lum}
\end{equation}
and use the numerical factor $\eta_a \leq 1$ as a measure
of the efficiency with which the accreted material is actually
converted into luminosity. Thus $\eta_a = 1$ for steady shell
burning, whereas $\eta_a = L_{\rm acc}/L_{\rm sb} \ltappeq 0.3$ for 
accretion onto a massive WD in the absence of any nuclear processing 
($L_{\rm acc} = GM_1\dot{M}_{\rm acc}/R_{1}$).
%\nocite{nauenberg1} 
Combining Equations~\ref{mdot_wind1} and~\ref{lum} yields 
\begin{equation}
\begin{array}{l} 
\dot{M}_{\rm acc} = - g \dot{M}_{w2} \\ 
\hspace*{0.9cm}
\simeq 1.2 \times 10^{-6} g^2 \phi^2 
\eta_a \eta_s \left(\frac{q^{5/2}}{1+q}\right)^{2/3} m_1 
{\rm ~M_{\sun}~yr^{-1}}.
\end{array}
\label{steady_mdotacc}
\end{equation}
Here, $g= -\dot{M}_{\rm acc}/\dot{M}_{w2}$ is the dimensionless accretion 
rate (measured in units of the wind mass-loss rate),
$m_1=M_1/M_{\sun}$, and we have used Paczy\'{n}ski's (1971)
approximation for $R_{L2}/a$ to recast $r_2/a_{11}$ in terms of
$q$. Here, $R_{L2}$ denotes the volume-averaged Roche-lobe radius of
the secondary. \nocite{paczynski3,iben1} If a stationary, stable,
wind-driven state exists, the accretion  rate in it must be given by
Equation~\ref{steady_mdotacc}.

In order to determine $g$ as a function of the system parameters, 
one needs to calculate the mass-transfer rate due to Roche-lobe
overflow (RLOF) in a semi-detached close binary system whose evolution
is driven by a stellar wind from the secondary. This calculation is 
described in detail by VK98 and KV98. Briefly, such a system will
quickly settle in a stationary state with $\dot{R_{L2}}/R_{L2} = 
\dot{R_2}/{R_2}$, provided such a state exists and is stable. The 
stationarity condition is sufficient to find $g$ as
\begin{equation}
g = \frac{(6\beta_2+2q)-(5+3\zeta)(1+q)}{(1+q)(5+3\zeta-6q)}, 
\label{g}
\end{equation}
where $\zeta$ is the effective mass-radius index of the secondary
(describing its reaction upon mass loss), and $\beta_2$ is the
specific angular momentum of the escaping wind material, measured
relative to the specific orbital angular momentum of the
secondary. 

Equation~\ref{g} gives the dimensionless RLOF mass-transfer
rate in the stationary wind-driven state. It is identical to
Equation~30 of VK98, except that we have retained the explicit
dependence on $\beta_2$ (VK98 set $\beta_2=1$ throughout). As shown by 
KV98, the stationary solution is stable for system parameters 
appropriate to T~Pyx and $\dot{M}_{w2} \propto L^{1/2} \propto
\dot{M}_{\rm acc}^{1/2}$. In deriving Equation~\ref{g}, it has been
assumed that all of the material in the stellar wind from 
the secondary escapes.
%and is not accreted by the primary. 
Also, the effects of mass loss from the primary have been ignored. In
reality, T~Pyx may undergo significant mass ejection during its nova 
eruptions, and the long term average $\dot{M}_{w1}$ may be comparable
to $\dot{M}_{w2}$. However, we have verified that even strong,
episodic mass loss from the primary has virtually no impact on our
results for T~Pyx in quiescence, unless the specific angular momentum
carried away by the nova ejecta is extremely high (much higher than
that of the primary). In that case, the angular momentum loss
associated with nova eruptions would further accelerate (and could
possibly even dominate) the binary evolution. We will return to this 
possibility in Section~\ref{conclusions}.

We finally consider the orbital period derivative of such a 
wind-driven system. If this is entirely due to {\em stationary} mass
loss and/or 
transfer, we can combine Kepler's 3rd law with
Paczy\'{n}ski's (1971) approximation for $R_{L2}$ and differentiate
logarithmically to obtain 
\begin{equation}
\frac{\dot{P}_{\rm orb}}{P_{\rm orb}} =
\frac{3\zeta-1}{2}\frac{\dot{M}_{2}}{M_2} =
\frac{(3\zeta-1)(1+g)}{2}\frac{\dot{M}_{w2}}{M_2}
\label{pdot}
\end{equation}
(c.f. Equation~34 in VK98).
However, the period derivative measured in real systems may differ
from this, since the timescale associated with such 
measurements is much shorter than the timescale on which the RLOF rate 
can adjust itself ($\tau_{\rm RLOF} \sim
\frac{H}{R_2}\frac{M_2}{\dot{M}_{w2}} \sim 10^4$~yr, where $H/R_2 \sim
10^{-4}$ is the scale height in the atmosphere of the secondary
near the inner Lagrangian point). Thus observationally determined
period changes could, for example, be due to fluctuations of 
$\dot{M}_{w2}$ or $\beta_2$ on timescales shorter than $\tau_{\rm RLOF}$.

\subsection{Application to T~Pyx}
\label{discussion1}

We are now ready to apply the wind-driven evolution scenario to
T~Pyx. For definiteness in our numerical estimates, we will adopt
main-sequence-based values for the mass and  radius of 
T~Pyx's secondary: $M_2 = 0.12~M_{\sun}$, $R_2 = 0.17~R_{\sun}$. 
These follow from Kepler's law, the orbital period, Patterson's (1998)
power law approximation to the the M-dwarf mass-radius relationship of
Clemens et al. (1998) and 
Paczy\'{n}ski's (1971) approximation for $R_{L2} \simeq R_{2}$ of a 
Roche-lobe filling secondary star.\nocite{paczynski3}   
In addition, we will use $M_1=1.2~M_{\sun}$ as an estimate of the 
white dwarf mass (Contini \& Prialnik 1997). We therefore take the 
mass ratio in T~Pyx to be $q=M_2/M_1=0.1$. We note from the outset
that even this rather low value for $q$ could be an overestimate,
since T~Pyx's wind-driven evolution may already have reduced $M_2$
below its main-sequence value. Our assumption that the secondary is
still close to the main sequence amounts to saying that wind-driving
has only just begun.\nocite{clemens1}

We begin by noting that, for a low-mass secondary undergoing adiabatic
mass loss, we may take $\zeta \simeq -1/3$ (VK98). Next, we consider
two extreme estimates for the angular momentum loss parameter 
$\beta_2$. To obtain a lower limit, we note that the stellar wind
material will carry away at least the specific angular momentum of the
secondary, in which case $\beta_2 = 1$ (this is the case considered by
VK98). On 
the other hand, the stellar wind material may extract angular momentum
from the binary system by frictional processes. An upper
limit to the amount of specific angular momentum that is likely to be
extracted this way is given by 
the specific angular momentum of particles escaping
through the outer Lagrangian points, which is $j_2 \simeq 1.65 a^2 \Omega$
(Sawada et al. 1984; Flannery \& Ulrich 1977; Nariai
1975). Here $\Omega$ is the angular velocity of the binary system,
and, in our notation, the corresponding angular momentum loss parameter is $\beta_2 = 2$. These estimates, together with $q=0.1$, yield 
$g \simeq 0.5$ ($\beta_2 = 1$) and $g \simeq 2$ ($\beta_2 = 2$). 
Substituting these values back into Equation~\ref{steady_mdotacc},
along with $m_1=1.2$, we obtain
\begin{equation} 
\dot{M}_{\rm acc}({\rm T~Pyx}) \simeq \left\{ \begin{array}{ll} 
7 \times 10^{-9} \phi^2 \eta_s \eta_a ~{\rm {M}_{\sun}~yr^{-1}} &
~~\beta_2 = 1 \\ \\
1 \times 10^{-7} \phi^2 \eta_s \eta_a ~{\rm {M}_{\sun}~yr^{-1}} &
~~\beta_2 = 2 
\end{array} \right.
\label{mdot_acc_tp}
\end{equation}
for T~Pyx if it has settled into a stationary, wind-driven state. The 
corresponding luminosity is given by Equation~\ref{lum} as
\begin{equation}
L({\rm T~Pyx}) \simeq  \left\{ \begin{array}{ll} 
2 \times 10^{36} \phi^2 \eta_s \eta_a^2 ~{\rm erg~s^{-1}} &
~~~~~~~\beta_2 = 1 \\ \\
3 \times 10^{37} \phi^2 \eta_s \eta_a^2 ~{\rm erg~s^{-1}} &
~~~~~~~\beta_2 = 2 ~ .
\end{array} \right.
\label{lum2}
\end{equation}
Thus wind-driving can indeed account for T~Pyx's extreme accretion
rate and luminosity, provided that the various efficiency factors in
Equations~\ref{mdot_acc_tp} and~\ref{lum2} are not too far from
unity. 

VK98 have argued that $\eta_s \simeq 1$ in SSSs, since soft
x-rays are absorbed well above the photosphere and should therefore be
quite efficient at driving the wind. The factor $\phi$ is just 
the area of the mass-losing regions on the secondary divided by $\pi
R_2^2$, so that mass loss from the entire front hemisphere would
correspond to $\phi = 2$. We may therefore also expect $\phi \simeq 1$,
even if the secondary is partially shielded by an optically thick
accretion disk. However, the most interesting parameter in this
context is $\eta_a$. As noted in Section~\ref{basics}, energy release by
accretion yields $\eta_a \ltappeq 0.3$. The upper limit in
this inequality corresponds to a Carbon-core WD of maximum mass
($1.4$~M$_{\odot}$) and minimum radius ($0.002R_{\odot}$; 
Hamada-Salpeter 1961). For a 1.2~$M_{\odot}$ WD on the same
mass-radius relation, we have $\eta \simeq 0.1$. Thus gravitational
energy release alone can only meet the system's luminosity
requirements if the WD is even more massive than we have assumed and
$\beta_2 > 1$.  

The alternative is that nuclear processing continues in T~Pyx even in
quiescence. This would imply that nuclear burning in T~Pyx occurs both
quasi-steadily (in quiescence) and explosively (during outbursts),
with the former taking place at a rate slightly below the accretion rate.
From an empirical point of view, this does not seem unreasonable: as
noted by Patterson et al. (1998), some SSSs in the LMC and M31 are
recurrent (see Kahabka 1995), and the galactic novae 
GQ~Mus and V1974 remained luminous soft x-ray sources for several
years after their nova eruptions (Shanley et al. 1995; Krautter et
al. 1996). Symbiotic stars (SySs) provide another interesting point of
comparison. Several classical SySs, such as Z~And, exhibit 
erupting behaviour even though though their quiescent luminosities ($L
\sim 10^3 L_{\odot}$) suggest nuclear processing as the dominant power
source (M\"{u}rset, Nussbaumer, Schmid \& Vogel 1991). 
In addition, the luminosities of symbiotic novae (a distinct
class from the classical SySs) remain high for decades after their
outbursts (M\"{u}rset \& Nussbaumer 1994).

From a theoretical point of view, the situation is somewhat more
difficult. The problem is that 
the rate at which steady nuclear processing can proceed is not, in
principle, a tunable parameter. Based on 1-dimensional models,
explosive and steady processing are generally expected to be mutually
exclusive regimes, whose dividing line is a function of accretion rate
and white dwarf mass (Iben 1982). Thus steady burning is not expected
to occur for accretion rates less than $\dot{M}_{\rm crit} \sim 1.32
\times 10^{-7} M_{\rm WD}^{3.57}$. The accretion rates predicted by our
wind-driven model are below this line (albeit by only a factor of 2.5
for $\beta =2$ and $M_{\rm WD} = 1.2 M_{\odot}$).On the other hand, nuclear
processing on real WDs is unlikely to be well-described by
spherically symmetric models. For example, thermonuclear runaways
(TNRs) triggered by localized temperature or pressure fluctuations
may not always spread and evolve into global TNRs (Shara 1982; Shankar 
\& Arnett 1994). Quasi-steady quiescent burning in 
T~Pyx could therefore conceivably be the collective outcome of many
successive localized TNRs. The individual mini-eruptions would not
necessarily be obvious observationally, provided they are sufficiently
small and frequent. These localised TNRs could process non-degenerate
material more slowly than it accretes, leading to the build-up of a
more and more massive non-degenerate envelope and, eventually, to a
global TNR.

The preceding is a highly speculative scenario, and we do not mean to
endorse it too strongly. We have outlined it mainly to provide a
specific illustration of the general idea that (quasi-)steady and
explosive nuclear processing might take place in the same object. This
general idea is not new. It was first suggested by Webbink et
al. (1986) and again by Patterson et al. (1998), both times without any
specific model in mind. The main achievement of the wind-driving mechanism 
is that, given a high radiative efficiency, T~Pyx's high mass
accretion rate and luminosity can be accounted for self-consistently.
We do, of course, acknowledge that the requirement $\eta_a \sim 1$ 
implies a fair amount of fine-tuning, regardless of whether one
invokes an extremely massive WD or
quiescent nuclear burning. But some theoretical fine-tuning seems
reasonable for a system like T~Pyx, whose short orbital period, high
luminosity and ability to produce nova eruptions are a unique
combination among CVs and SSSs. This is not to say that the
wind-driven evolutionary channel must be narrow: most other
wind-driven systems may be characterized by somewhat higher accretion
rates than T~Pyx and may therefore be {\em steady}
SSSs. Observationally, such steady, wind-driven SSSs may nevertheless
be rare, since their evolutionary timescales would be even shorter
than T Pyx's. 

We finally turn to the orbital period derivative. On substituting the 
values for $g$ and $\dot{M}_{w2} = -g \dot{M}_{\rm acc}$ derived above
into Equation~8 and inverting, we find that the expected timescale for 
period increase due to {\em stationary} wind-driven mass loss and mass
transfer in T~Pyx is 
\begin{equation}
\frac{P_{\rm orb}}{\dot{P}_{\rm orb}}\left({\rm T~Pyx}\right)
 \simeq  \left\{ \begin{array}{ll} 
5\times 10^6 ~~{\rm yr} & ~~~~~~~~~~~\beta_2 = 1 \\ \\
7\times 10^5 ~~{\rm yr} & ~~~~~~~~~~~\beta_2 = 2 ~ .
\end{array} \right. 
\label{pdot2}
\end{equation}
Patterson et al. (1998) derived $P_{\rm orb}/\dot{P}_{\rm orb} \simeq 3 \times
10^5$~yr from the periodic dip in T~Pyx's optical light curve, but
more recent timings suggest a slower rate of change. In any case, as
noted in the discussion following Equation~\ref{pdot}, the observed
$\dot{P}_{\rm orb}$ may not be a valid estimate of the stationary
(long-term average) value.

\section{Discussion and Conclusions}
\label{conclusions}

We have shown that the abnormally high luminosity of the 
short-period recurrent nova T~Pyx can be explained if the system
belongs to the class of wind-driven SSSs. T~Pyx's accretion rate is
then much higher than that of an ordinary CV, because the system's
evolution is dominated by a radiation-induced wind from the secondary
star. The wind-driven evolution is self-sustaining, because the
wind-induced accretion rate is sufficient to power the luminosity that
excites the wind.

Our model predicts that T~Pyx is an intrinsically bright soft
x-ray source. T~Pyx was observed serendipitously as part of the ROSAT 
all-sky survey, but not detected (Verbunt et al. 1997). This is 
not yet in conflict with a SSS model for T~Pyx, because of the limited
sensitivity of these observations and the relatively high column
density towards the system ($N_{\rm H} \simeq 2.1 \times 10^{21}$~cm$^{-2}$;
Dickey \& Lockman 1990\footnote{Data available online at
{\em http://heasarc.gsfc.nasa.gov/cgi-bin/Tools/w3nh/w3nh.pl}.}).
However, a sensitive, pointed observation with Chandra or XMM could 
yield a detection and provide a direct estimate of the WD luminosity 
and temperature. For example, assuming a distance of 3~kpc
(e.g. Patterson et al. 1998), the flux produced by a blackbody with
$L_{\rm WD} = 10^{37}$~ergs/s and $T_{\rm WD}=2.4\times 10^5$~K is below the
ROSAT detection limit, but would provide around 1000 counts in a
30~ksec observation with XMM. 
\nocite{dickey1}\nocite{verbunt1}

If T~Pyx is, in fact, a wind-driven SSS, its current evolutionary
timescale is $\tau_{\rm evol} = |M_2/\dot{M}_{2}| = \dot{P}_{\rm orb}/P_{\rm orb}
\sim 10^{6}$~yrs. Clearly, this is an extremely short-lived evolutionary 
state. So how will T~Pyx's evolution actually end? We can think of at
least three possible outcomes. First, the mass ratio of the system 
may be driven so low that irradiation and wind-driving cease to
be effective. Second, the wind-driven evolution may proceed until the 
secondary completely evaporates, leaving T~Pyx as an isolated
WD. Third, if the total mass of the system is somewhat higher than 
we have assumed, the WD may reach the Chandrasekhar limit and explode
in a Type~Ia supernova. 

The likelihood of the last outcome depends on whether the amount of
material that is ejected in each of T~Pyx's recurrent nova outbursts
is significantly less than that which is accreted during each
quiescent interval. Theoretically, this is plausible 
(e.g. Prialnik \& Kovetz 1995; Livio \& Truran 1992). The question
might also be addressed observationally: the total mass accreted during
quiescence is $\Delta M_{\rm acc} \simeq \dot{M}_{\rm acc} \tau_{\rm rec} \simeq
1-20 \times 10^{-7}~{\rm M_{\sun}}$. The mass ejected during an outburst 
can be inferred by comparing the pre- and post-outburst orbital
periods, although this is sensitive to the details of the ejection 
process (Livio~1991; Livio et al. 1991). Such a comparison could also 
distinguish between wind-driven and nova-driven evolution
scenarios. As noted in Section~\ref{basics}, if the specific angular
momentum of the mass ejected during T~Pyx's nova eruptions is very
high ($j_{\rm nova} \gtappeq a^2\Omega >> j_1$), the angular momentum loss
associated with nova eruptions could, in principle, itself drive
T~Pyx's high mass transfer rate. In the context of this model, the
difference between pre- and post-outburst orbital periods must be
relatively large. By contrast, the wind-driving model makes no such
prediction.
\nocite{livio6,livio7} 

Given the strong possibility that the system will destroy itself on a
short timescale, T~Pyx represents an evolutionary channel by which
short-period CVs may be 
removed from the general CV population. The existence of such a
channel is desirable from a theoretical point of view. As most
recently pointed out by Patterson (1998), the observed CV population
shows a significant dearth of 
short-period systems relative to the predictions of standard 
evolutionary scenarios. Whether T~Pyx's "assisted stellar suicide"
represents a sufficiently wide channel in this context is an important
question: do most CVs eventually enter a
wind-driven phase or is T~Pyx really a unique system? The answer
clearly depends on how wind-driving is actually 
triggered, an issue we have not addressed in the present work. 
One possibility is that triggering
occurs during a  period of residual nuclear burning in the aftermath
of a nova eruption, though this would have to be limited to systems
satisfying other constraints (e.g. low $q$ and/or high $M_{\rm WD}$).
%In our {\em stationary}, wind-driven model for T~Pyx, we have simply 
%assumed that triggering has, in fact, taken place. We are now in the
%process of identifying and investigating the most likely candidate 
%mechanisms. 

\nocite{patterson2}

\begin{acknowledgements}
We are grateful to Jim Truran for a clear, concise and cogent
referee's report. Support for CK was provided by NASA through Hubble
Fellowship grant HF-01109 awarded by the Space Telescope Science 
Institute, which is operated by the Association of Universities for
Research in Astronomy, Inc., for NASA under contract NAS 5-26555.
ARK gratefully acknowledges a PPARC Senior Fellowship.
\end{acknowledgements}

\bibliographystyle{apj}
\bibliography{bibliography}
\end{document}